\documentclass[aps,graphicx,6pt,showkeys,showpacs,onecolumn]{revtex4}
\usepackage{amsmath}
\usepackage{amscd}
\usepackage{graphicx}
\usepackage{subfigure}
\usepackage{appendix}
\usepackage{amsfonts}
\usepackage{multirow}

\begin{document}

\title[Short Title]{Arbitrary quantum state engineering in three-state systems via Counterdiabatic driving}

\author{Ye-Hong Chen$^{1}$}
\author{Yan Xia$^{1,}$\footnote{E-mail: xia-208@163.com}}
\author{Qi-Cheng Wu$^{1}$}
\author{Bi-Hua Huang$^{1}$}
\author{Jie Song$^{2,}$\footnote{E-mail: jsong@hit.edu.cn}}

\affiliation{$^{1}$Department of Physics, Fuzhou University, Fuzhou 350002, China\\
             $^{2}$Department of Physics, Harbin Institute of Technology, Harbin 150001, China}


\begin{abstract}
  A scheme for arbitrary quantum state engineering (QSE) in three-state systems is proposed.
  Firstly, starting from a set of complete orthogonal time-dependent basis
  with undetermined coefficients, a time-dependent Hamiltonian is derived via Counterdiabatic driving for the purpose of
  guiding the system to attain an arbitrary target state at a predefined time. Then, on request of the assumed target states,
  two single-mode driving protocols and a multi-mode driving protocol are proposed as examples to discuss the validity of the QSE scheme.
  The result of comparison between single-mode driving and multi-mode driving
  shows that multi-mode driving seems to have a wider rang of application prospect
  because it can drive the system to an arbitrary target state from an arbitrary initial state also at a predefined time
  even without the use of microwave fields for the transition between
  the two ground states. Moreover, for the purpose of discussion in the scheme's feasibility in practice,
  a polynomial ansatz as the simplest exampleis used to fix the pulses.
  The result shows that the pulses designed to implement the protocols are not hard to be realized in practice.
  At the end, QSE in higher-dimensional
  systems is also discussed
  in brief as a generalization example of the scheme.
\end{abstract}


\maketitle
\section{Introduction}

In recent years, to fulfill the requirement of high-precision quantum gates, teleportation, or state transfer,
much focus is given to quantum state engineering (QSE)
\cite{Pla26494,Sci292472,Sci2921695,Prl90127902,Prl103120504,Prl109115703,Pra87022332,Jpa42365303,Prl104063002,Pra83062116,Njp14093040,Pra88033406,Pla3753343}
which aims to manipulate the system and
obtain a target state, typically a pure state, at a designed time $T$,
or more ambitiously, to drive the eigenstates of an initial
Hamiltonian into those of a final Hamiltonian \cite{Jpa42365303,Prl104063002,Pra83062116,Njp14093040,Pra88033406}. To be
concrete about it, taking two-level system as an example, the goal of QSE is to construct a passage to achieve
an expected final state $|\psi(T)\rangle=\mu|1\rangle+\nu|2\rangle$ ($|\mu|^{2}+|\nu|^{2}=1$) from a
given initial state $|\psi(0)\rangle$ with a designed evolution time $T$ in an undisturbed way.
Typically, adiabatic techniques behave very well in the field of QSE.
In an adiabatic control of a quantum system, the system remains in one of the instantaneous eigenstates of its time-dependent
Hamiltonian during the entire evolution. The control parameters in the Hamiltonian are carefully designed such that the
adiabaticity condition always holds, which usually results in a very long execution time.
By adiabatic passages and sequential programming \cite{Pla26494,Sci292472,Sci2921695,Prl90127902,Prl103120504,Prl109115703,Pra87022332,Rmp701003,Rmp7953},
robust protocols \cite{Pra82030303,Pra88052326} of realizing QSE have been provided in closed-system scenario.
As the system remains in the instantaneous eigenstates,
there is no heating or friction, but the long operation times needed may render the operation useless or even impossible to implement
because decoherence would spoil the intended dynamics. Therefore,
accelerating the dynamics towards the perfect final outcome is
a good idea and perhaps the most reasonable way to actually
fight against the decoherence that is accumulated during a long operation time.
In this field, ways of speeding up an adiabatic control are available in the adiabatic regime \cite{Jpa42365303,Prl100113601,Jpb42241001,Pra93052109,Amp62117,Prl105123003},
for instance, in ref. \cite{Prl105123003}, based on Berry's transitionless tracking algorithm \cite{Jpa42365303}, Chen \emph{et al.}
put forward a shortcut to adiabatic passage in two- and three-level systems. Soon after that, lots of speeding up protocols have been springing up
and have been applied in a wide range of fields including ``fast coldatom'',
``fast ion transport'', ``fast expansions'', ``fast wave-packet splitting'', ``fast quantum
information processing'', and so on \cite{Pra86033405,Pra13013415,Pra85033605,
Pra89033403,Pra84043434,Pra89012326,Pra84031606Epl9660005,Pra84023415,CYH,Pra87043402,Pra89043408,Pra89053408,Prl109100403,Pra82053403,Njp13113017,Pra88053422}.

The transitionless tracking algorithm which is also known as Counterdiabatic driving, provides
Hamiltonians $H(t)$ for which the adiabatic approximation
for the time-dependent wave function evolving
with a reference Hamiltonian $H_{0}(t)$ becomes exact.
According to ref. \cite{Jpa42365303}, the simplest Hamiltonian which steers
the dynamics along the instantaneous eigenstates $\{|\phi_{n}(t)\rangle\}$ of the original Hamiltonian $H_{0}(t)$ without
transitions among them and without phase factors, formally in an
arbitrarily short time, $H(t)=i\hbar\sum_{n}|\dot{\phi}_{n}(t)\rangle\langle\phi_{n}(t)|$, where the dot means time derivative.
Strictly speaking, the form of the Hamiltonian deduced by transitionless tracking algorithm is
$H(t)=H_{0}(t)+H_{1}(t)$, where $H_{1}(t)=i\hbar\sum_{n}|\dot{\phi}_{n}(t)\rangle\langle\phi_{n}(t)|$.
While, in fact, the addition of $H_{0}(t)$ is possible, but not necessary, it only affects the phases \cite{Prl105123003}.
So $H_{1}$ may substitute $H_{0}(t)$, when $H(t)=H_{1}(t)$. In this case, the evolution operator of the system could be described as $U(t)=\sum_{n}|\phi_{n}(t)\rangle\langle\phi_{n}(\tau_{0})|$.

We should notice that if we pay no attention to that $\{|\phi_{n}(t)\rangle\}$ are the instantaneous eigenstates of $H_{0}(t)$,
the Hamiltonian $H(t)$ seems to be irrelevant to $H_{0}(t)$ but only closely related to $|\phi_{n}(t)\rangle$.
Therefore, assuming $\{|\phi_{n}(t)\rangle\}$ are just a set of complete orthogonal basis that $\{|\phi_{n}(t)\rangle\}$ satisfy
$\sum_{n}|\phi_{n}(t)\rangle\langle\phi_{n}(t)|=1$ and $\langle\phi_{n}(t)|\phi_{m}(t)\rangle=\delta_{nm}$.
According to transitionless tracking algorithm, when the Hamiltonian for driving the system is $H(t)=i\hbar\sum_{n}|\dot{\phi}_{n}(t)\rangle\langle\phi_{n}(t)|$,
each of the moving states $|\phi_{n}(t)\rangle$ will evolve along itself all the time without transition to others.
In other words, $\{|\phi_{n}(t)\rangle\}$ are not necessarily being the instantaneous eigenstates of an original Hamiltonian $H_{0}(t)$, as long as they
satisfy orthonormality, the corresponding Counterdiabatic driving Hamiltonian $H(t)$ could be deduced.
In this case, suitable pathes can be designed for different purposes as one desired.
With the designed paths, full information including the populations of the states and the phases at any time would be
exactly known.  Here, it is important to note that
shortening the time implies an energy cost, so $T$ could be arbitrarily value only when the energy cost could be arbitrarily value. Which means the
advantage of obtaining an arbitrarily target state with an arbitrarily operation time $T$ is constrained only the complementarity energy-time
\cite{Sr515775,Pra93012311,arXiv1601,arXiv1602,Njp18023001}.

In this paper, motivated by refs.
\cite{Jpa42365303,Prl105123003,Pra88053422}, we start from a set of
undetermined moving states which satisfy orthonormality to deduce a
Hamiltonian based on transitionless tracking algorithm to implement
arbitrary QSE scheme in three-state systems. We first
consider the coefficients $\mu$, $\eta$, and $\nu$ of the expected
state $|\psi(T)\rangle=\mu|1\rangle+\eta|2\rangle+\nu|3\rangle$ are
all real. Then, according to transitionless tracking algorithm, we
deduce the corresponding Counterdiabatic driving Hamiltonian. The
third step, through using the initial state, the final states, and
the limiting conditions for a realizable Hamiltonian, we set the
boundary conditions. At last, the pulses (or the coupling
coefficients) are determined, which means the Hamiltonian will be
constructed. And the constructed Hamiltonian will accurately guide
the system from the given initial state to the expected state with a
designed evolution time $T$. To show the work in more detail, two single-mode driving protocols are proposed, and the result
of numerical analysis show that, as expected, the target state could
be ideally achieved along the passage constructed. Moreover, a protocol of multi-mode driving is proposed later. We show that even
without using the microwave fields for the transition between the
two ground states (the 1-3 pulse), the multi-mode driving protocol
still can drive the system to an arbitrary target state from an
arbitrary initial state, which makes sense in application prospect.
We give an example to discuss the situation when the
coefficients $\mu$, $\eta$, and $\nu$ contain complex phases.
Finally, the generalization of higher-dimensional systems and
several application examples of the present QSE scheme are given.


\section*{QSE in three-state systems via counterdiabatic driving}
We start from constructing a complete orthogonal basis for a three-state system with three bare states $|1\rangle$, $|2\rangle$, and $|3\rangle$,
\begin{eqnarray}\label{eq1-1}
  |\phi_{n}(t)\rangle=\cos{\alpha_{n}}\cos{\beta_{n}}|1\rangle+\sin{\beta_{n}}|2\rangle+\sin{\alpha_{n}}\cos{\beta_{n}}|3\rangle,
\end{eqnarray}
where $\alpha_{n}$ and $\beta_{n}$ are time-dependent real coefficients.
To satisfy the orthogonality condition that $\langle\phi_{n}(t)|\phi_{m}(t)\rangle=0$ ($n\neq m$), we find $\alpha_{n}$ and $\beta_{n}$ satisfy the condition
\begin{eqnarray}\label{eq1-2}
  \cos(\alpha_{n}-\alpha_{m})\cos{\beta_{n}}\cos{\beta_{m}}+\sin{\beta_{n}}\sin{\beta_{m}}=0.
\end{eqnarray}
Then, according to transitionless tracking algorithm \cite{Jpa42365303}, the Hamiltonian that exactly
drives the moving states is derived in the form ($\hbar=1$),
\begin{eqnarray}\label{eq1-3}
  H(t)=i\sum_{n=1}^{3}{|\dot{\phi}_{n}(t)\rangle\langle\phi_{n}(t)|}.
\end{eqnarray}
To satisfy the condition in eq. (2), the simplest choice for the coefficients $\alpha_{n}$ and $\beta_{n}$ could be
\begin{eqnarray}\label{eq2-1}
  \alpha_{1}-\alpha_{2}=\alpha_{1}-\alpha_{3}=\beta_{2}-\beta_{3}=\frac{\pi}{2}, \ \ \beta_{1}=0.
\end{eqnarray}
For convenience, we set $\alpha_{1}=\theta$ and $\beta_{2}=\varphi$. Then,
the three moving states become $|\phi_{1}(t)\rangle=\cos{\theta}|1\rangle+\sin{\theta}|3\rangle$,
$|\phi_{2}(t)\rangle=-\sin{\theta}\cos{\varphi}|1\rangle-\sin{\varphi}|2\rangle+\cos{\theta}\cos{\varphi}|3\rangle$,
and
$|\phi_{3}(t)\rangle=-\sin{\theta}\sin{\varphi}|1\rangle+\cos{\varphi}|2\rangle+\cos{\theta}\sin{\varphi}|3\rangle$.
Putting eq. (\ref{eq2-1}) into eq. (\ref{eq1-3}), the Hamiltonian is deduced
\begin{eqnarray}\label{eq2-2}
  H(t)&=&i(-\dot{\varphi}\sin\theta|1\rangle\langle2|+\dot{\varphi}\sin\theta|2\rangle\langle1|\cr\cr
                        &&-\dot{\varphi}\cos{\theta}|2\rangle\langle3|+\dot{\varphi}\cos{\theta}|3\rangle\langle2|\cr\cr
                        &&-\dot{\theta}|1\rangle\langle3|+\dot{\theta}|3\rangle\langle1|).
\end{eqnarray}
So, supposing that the atomic system is $\Lambda$-type for the following discussion
(two ground states $|1\rangle$, $|3\rangle$, one excited state $|2\rangle$).
Accordingly, we set $\Omega_{p}(t)=\dot{\varphi}\sin{\theta}$, $\Omega_{s}(t)=\dot{\varphi}\cos{\theta}$, and $\Omega_{a}(t)=-\dot{\theta}$,
where $\Omega_{p}(t)$, $\Omega_{s}(t)$, and $\Omega_{a}(t)$ can be regarded as the pump, Stokes, and microwave fields, respectively.
The aim is to obtain a fast population transfer 
to create an arbitrary stable superposition state $|\psi(T)\rangle=\mu|1\rangle+\eta|2\rangle+\nu|3\rangle$ ($|\mu|^{2}+|\eta|^{2}+|\nu|^{2}=1$)
from a given initial state, i.e., $|1\rangle$, along a chosen moving state $|\phi_{k}(t)\rangle$ ($k=1$, or $2$, or $3$).
Therefore, according to eq. (\ref{eq1-1}), the boundary condition is
\begin{eqnarray}\label{eq1-6}
   \beta_{k}(\tau_{0})&=&0,\ \alpha_{k}(\tau_{0})=0,   \cr\cr
   \beta_{k}(\tau_{f})&=&\arcsin{\eta}, \ \alpha_{k}(\tau_{f})=\arctan(-\frac{\mu}{\nu}),
\end{eqnarray}
where $\tau_{0}$ is the initial time, $\tau_{f}$ is the final time, and $T=\tau_{f}-\tau_{0}$ is the total operation time.
Based on the Hamiltonian, anyone of the three moving states $\{|\phi_{n}(t)\rangle\}$ could be chosen to complete the QSE scheme. Now, we are ready to apply
QSE by means of different protocols including single-mode driving and multi-mode driving.
Here, the single-mode driving denotes the system accurately evolves alone one of the moving states.
In other words, when the system is initially in one of the three moving states, $\langle\psi(\tau_{0})|\phi_{k}(\tau_{0})\rangle=1$ ($k=1$, or 2, or 3),
it will be in that moving state all the time, $\langle\psi(t)|\phi_{k}(t)\rangle=1$, where $|\psi(t)\rangle$
is the evolution state given by solving the Schr\"{o}dinger equation $i\frac{\partial}{\partial t}|\psi(t)\rangle=H(t)|\psi(t)\rangle$.
The multi-mode driving means that the time-dependent wave function
$|\psi(t)\rangle$ will include contributions from the three moving states.
That is, when the system is initially in a linear superposition of the moving states,  $|\psi(\tau_{0})\rangle=\sum_{k}c_{k}|\phi_{k}(\tau_{0})\rangle$
($c_{k}$ is a time-independent coefficient satisfying $\sum_{k}|c_{k}|^{2}=1$), at the time $t$, the system will be also
in a linear superposition of the moving states $|\psi(t)\rangle=\sum_{k}c_{k}|\phi_{k}(t)\rangle$.

\section*{Single-mode driving}
\subsection*{Protocol I}
In the first single-mode driving protocol,  assuming that the aim now is to obtain an arbitrary stable superposition
state $|\psi(T)\rangle=\mu|1\rangle+\nu|3\rangle$ ($|\mu|^2+|\nu|^2=1$)
between the two ground states $|1\rangle$ and $|3\rangle$ in a classical $\Lambda$-type atom.
Then, $|\phi_{1}(t)\rangle$ could be chosen to gain the target.
Since the moving state $|\phi_{1}(t)\rangle$ is irrelevant to $\varphi$, so in fact we can
further simplify the Hamiltonian in eq. (\ref{eq2-2}) by setting $\varphi=$const. $H(t)$ becomes $H(t)=-i\dot{\theta}|1\rangle\langle3|+H.c.$, and the
boundary condition is (for simplicity, we set $\tau_{0}=0$ and $\tau_{f}=T$)
\begin{eqnarray}\label{eq2-3}
&&\theta(0)=0, \theta(T)=\arcsin{\nu}, \cr\cr
&&\dot{\theta}(0)=0, \dot{\theta}(T)=0.
\end{eqnarray}
To satisfy this boundary condition, the simplest choice could be assuming a polynomial ansatz to interpolate at intermediate time,
$\theta(t)=\sum_{j=0}^{3}a_{j}t^{j}$. Then, putting eq. (\ref{eq2-3}) into $\theta(t)$, we obtain
\begin{eqnarray}\label{eq2-4}
  a_{0}=a_{1}=0, \ a_{2}=\frac{3\arcsin{\nu}}{T^{2}},\ a_{3}=-\frac{2\arcsin{\nu}}{T^{3}}.
\end{eqnarray}
Therefore, $\dot{\theta}=6(\frac{t}{T^{2}}-\frac{t^{2}}{T^{3}})\arcsin{\nu}$.
Once $\dot{\theta}$ is fixed, which means $\Omega_{a}(t)$ is fixed,
we may calculate the time-evolution for pulse and population (see, e.g., Fig. 1, where $\mu=\nu=1/\sqrt{2}$ is chosen as an example).
Here the population for a quantum state $|\Psi\rangle$ is defined as $P=|\langle\Psi|\psi(t)\rangle|^{2}$, where $|\psi(t)\rangle$ is the solution of
the Schr\"{o}dinger equation $i\partial_{t}|\psi(t)\rangle=H|\psi(t)\rangle$.
From Fig. 1, two results could be easily got: the pulse for driving the system is easy to be realized in practice and
the system accurately evolve along the moving state $|\phi_{1}(t)\rangle$.
The pulse could be simulated by the sine curve algorithm, $\Omega_{a}=-\dot{\theta}\approx\Omega_{0}\sin(\pi t/T)$, where $\Omega_{0}$ is
the pulse amplitude. A classic application example of protocol I is the realization of an arbitrarily fast populations inversion between the ground states as shown in
Fig. 2 when $\nu=1$. The result shows that the population inversion could
be realized in an arbitrarily interaction time $T$.

It is worth noting that in confirming the boundary condition, there
would be two results for $\theta(T)$, say, $\theta(T)=\arccos{\mu}$
and $\theta(T)=\arcsin{\nu}$. Once the target state requests
$\mu<0$, for $\theta(T)=\arccos{\mu}$, there are two solutions:
$\theta(T)=\pi+\arccos{|\mu|}$ and $\theta(T)=\pi-\arccos{|\mu|}$,
where $\arccos{|\mu|}=\arcsin{|\nu|}$. Then, putting these two
solutions into
$|\psi(t)\rangle=\cos{\theta(t)}|1\rangle+\sin{\theta(t)}|3\rangle$,
for $\theta(T)=\pi+\arccos{|\mu|}$, we have
$|\psi(T)\rangle=-|\mu||1\rangle-|\nu||3\rangle$, while for
$\theta(T)=\pi-\arccos{|\mu|}$, we have
$|\psi(T)\rangle=-|\mu||1\rangle+|\nu||3\rangle$.
Meanwhile, when $\nu<0$, for $\theta(T)=\arcsin{\nu}$, the two
solutions are $\theta(T)=\pi+\arcsin{|\nu|}$ and
$\theta(T)=-\arcsin{|\nu|}$. These two solutions correspond to the
target states $|\psi(T)\rangle=-|\mu||1\rangle-|\nu||3\rangle$ and
$|\psi(T)\rangle=|\mu||1\rangle-|\nu||3\rangle$, respectively. In
order to embody the difference caused by the choices of $\theta(T)$,
we plot Fig. 3, which shows the fidelities $F_{n}$ of the
three bare states $|1\rangle$, $|2\rangle$, and $|3\rangle$ with
$|\mu|=|\nu|=1/\sqrt{2}$ as an example. We define $F_{n}=\langle
n|\psi(t)\rangle$ ($n=1,2,3$) as the fidelity for the state
$|n\rangle$ in plotting the Fig. 3. As shown in Fig. 3,
both the population and the phase that affecting the coefficients of
the system are evolved as expected. In fact, taking the global
phases off, $-|\mu||1\rangle-|\nu||3\rangle$ and
$-|\mu||1\rangle+|\nu||3\rangle$ are actually equivalent to
$|\mu||1\rangle+|\nu||3\rangle$ and $|\mu||1\rangle-|\nu||3\rangle$,
respectively. Therefore, when the global phase of the target state
is not in view, in order to reduce the pulse intensity to reduce the
energy consumption, $\theta(T)=\arcsin{\nu}$ would be the best
choice to confirm the boundary condition to implement the protocol
according to
$\dot{\theta}=6(\frac{t}{T^{2}}-\frac{t^{2}}{T^{3}})\theta(T)$ which
decides the pulse intensity.

\subsection*{Protocol II}
Now, we assume the aim is to obtain an arbitrary superposition state in
three levels $|\psi(T)\rangle=\mu|1\rangle+\eta|2\rangle-\nu|3\rangle$ ($\mu^{2}+\eta^{2}+\nu^{2}=1$).
We consider $\mu>0$, $\eta>0$, and $\nu>0$ as a matter of convenience for the discussion in the following.
According to eqs. (\ref{eq1-1}) and (\ref{eq1-6}),
we choose $|\phi_{2}(t)\rangle=\sin{\theta}\cos{\varphi}|1\rangle+\sin{\varphi}|2\rangle-\cos{\theta}\cos{\varphi}|3\rangle$
as the moving state.
The same as protocol I in Sec. III A, there also exist different choices for setting the boundary condition.
Nevertheless, since the difference only happens in the signs of the coefficients, to avoid the fussy, duplication and repetition,
we choose the boundary condition on the principle of less energy consumption (we set $\tau_{0}=0$ and $\tau_{f}=T$),
\begin{eqnarray}\label{eq2b-1}
  \dot{\varphi}(0)&=&0,\ \dot{\varphi}(T)=0,  \cr\cr
  \dot{\theta}(0)&=&0,\ \dot{\theta}(T)=0,     \cr\cr
  \varphi(0)&=&0,\ \varphi(T)=\arcsin{\eta}, \cr\cr
  \theta(0)&=&\frac{\pi}{2},\ \theta(T)=\arctan(\frac{\mu}{\nu}).
\end{eqnarray}
Similar to protocol I, we set $\theta=\sum_{j=0}^{3}a_{j}t^{j}$ and $\varphi=\sum_{j=0}^{3}b_{j}t^{j}$.
In this case, we obtain
\begin{eqnarray}\label{eq2b-2}
  a_{0}&=&\frac{\pi}{2},\ a_{1}=0,\ a_{2}=\frac{6\chi-\pi}{2T^{2}},\ a_{3}=-\frac{6\chi-3\pi}{3T^{3}}, \cr\cr
  b_{0}&=&0,\ b_{1}=0, \ b_{2}=\frac{3\arcsin{\eta}}{T^{2}},\ b_{3}=-\frac{2\arcsin{\eta}}{T^{3}},
\end{eqnarray}
where $\chi=\theta(T)=\arctan(\frac{\mu}{\nu})$.
Then $\theta$ and $\varphi$ are fixed,
\begin{eqnarray}\label{eq2b-3}
  \theta&=&\frac{\pi}{2}+\frac{(6\chi-3\pi)t^{2}}{2T^{2}}-\frac{(6\chi-3\pi)t^{3}}{3T^{3}}, \cr\cr
  \varphi&=&\frac{(3\arcsin{\eta})t^{2}}{T^{2}}-\frac{(2\arcsin{\eta})t^{3}}{T^{3}},    \cr\cr
  \dot{\theta}&=&\frac{(6\chi-3\pi)t}{T^{2}}-\frac{(6\chi-3\pi)t^{2}}{T^{3}},              \cr\cr
  \dot{\varphi}&=&\frac{(6\arcsin{\eta})t}{T^{2}}-\frac{(6\arcsin{\eta})t^{2}}{T^{3}}.
\end{eqnarray}
The Rabi frequencies for driving the system are
\begin{eqnarray}\label{eq2b-4}
  \Omega_{p}&=&\dot{\varphi}\sin{\theta}=(6\arcsin{\eta})(\frac{t}{T^{2}}-\frac{t^{2}}{T^{3}})\sin{\theta},\cr\cr
  \Omega_{s}&=&\dot{\varphi}\cos{\theta}=(6\arcsin{\eta})(\frac{t}{T^{2}}-\frac{t^{2}}{T^{3}})\cos{\theta}, \cr\cr
  \Omega_{a}&=&-\dot{\theta}=(3\pi-6\chi)(\frac{t}{T^{2}}-\frac{t^{2}}{T^{3}}).
\end{eqnarray}
In Fig. 4, we plot the evolutions in time of the Rabi frequencies with $\mu=0$, $\eta=\nu=1/\sqrt{2}$ and with $\mu=\eta=\nu=1/\sqrt{3}$ as examples.
We accordingly plot the evolutions in time of the populations with $\{\mu=0$, $\eta=\nu=1/\sqrt{2}\}$ and with $\{\mu=\eta=\nu=1/\sqrt{3}\}$ in Fig. 5.
The results show that the expected states can be ideally obtained with a given evolution time $T$ without doubt.
Here, it is notable that to realize the pulse with Rabi frequency $\Omega_{a}(t)$, in this protocol, we should
substitute the original two-photon transition in a stimulated Raman adiabatic
passage by a special one-photon $1-3$ pulse. This operation possibly can be
realized by exploiting the atomic clock transition between ground state hyperfine
levels ($|1\rangle$ and $|3\rangle$) with microwaves in alkali atoms \cite{Prl105123003}.

In fact, there is a special case that we can make the pulses more simple by suitably choosing parameters.
After reanalysing the moving state $|\phi_{2}\rangle$, we find  if the initial state is $|2\rangle$, through time evolution,
the moving state will also end up with an arbitrary stable superposition state $|\psi(T)\rangle=\mu|1\rangle+\eta|2\rangle-\nu|3\rangle$
with a time-independent special $\theta$.
In this case, $\Omega_{a}=0$, which means the microwave fields for the transition
between $|1\rangle$ and $|3\rangle$ to realize the special one-photon 1-3 pulse is no longer required and
the protocol maybe easier to be realized in practice.
We will discuss this special case in the following.

When $\theta$ is a const, the boundary condition should be a little different from eq. (\ref{eq2b-1}).
The initial state is no longer in state $|1\rangle$ but changes into $|2\rangle$. Then, the boundary condition becomes ($\tau_{0}=0$ and $\tau_{f}=T$)
\begin{eqnarray}\label{eq2b-5}
  \dot{\varphi}(0)&=&0,\ \dot{\varphi}(T)=0,  \cr\cr
  \dot{\theta}(0)&=&0,\ \dot{\theta}(T)=0,     \cr\cr
  \varphi(0)&=&\frac{\pi}{2},\ \varphi(T)=\arcsin{\eta}, \cr\cr
  \theta(0)&=&\theta(T)=\arctan{(\frac{\mu}{\nu})}.
\end{eqnarray}
The same as protocol I, we obtain the polynomial function for $\varphi$
\begin{eqnarray}\label{eq2b-6}
  \varphi&=&(\frac{3t^{2}}{T^{2}}-\frac{2t^{3}}{T^{3}})(\arcsin{\eta}-\frac{\pi}{2})+\frac{\pi}{2}, \cr\cr
  \dot{\varphi}&=&(\frac{6t}{T^{2}}-\frac{6t^{2}}{T^{3}})(\arcsin{\eta}-\frac{\pi}{2}),
\end{eqnarray}
and the Rabi frequencies
\begin{eqnarray}\label{eq2b-7}
  \Omega_{p}&=&\dot{\varphi}\sin{\theta}=6(\arcsin{\eta}-\frac{\pi}{2})(\frac{t}{T^{2}}-\frac{t^{2}}{T^{3}})\sin{\chi},\cr\cr
  \Omega_{s}&=&\dot{\varphi}\cos{\theta}=6(\arcsin{\eta}-\frac{\pi}{2})(\frac{t}{T^{2}}-\frac{t^{2}}{T^{3}})\cos{\chi}, \cr\cr
  \Omega_{a}&=&0.
\end{eqnarray}
In this case, the Rabi frequencies are shown in Fig. 6 with two set of parameters $\mu=\nu=1/\sqrt{2}$, $\eta=0$
and $\mu=1/\sqrt{6}$, $\eta=1/\sqrt{3}$, $\nu=1/\sqrt{2}$ as examples.
As we can see from Fig. 5, the shapes of the pulses are sinusoid.
Also, we accordingly deduce the populations versus time with these two set of pulses (see Fig. 7), which
demonstrates that the expected superposition states are ideally achieved at the designed time $t=T$.

\section*{Multi-mode driving}
The protocols proposed above in Sec. III are all based on single-mode driving which means only one of the moving states participates in the evolution.
In fact, since each of the moving states $|\phi_{n}(t)\rangle$ will evolve along itself all the time without transition to other
ones, multi-mode driving is also applicable for the QSE scheme. That is, the initial state is not necessary in one of the moving states,
it is feasible to set $|\psi(0)\rangle=\sum_{n}c_{n}|\phi_{n}(\tau_{0})\rangle$, where $c_{n}=\langle\phi_{n}(\tau_{0})|\psi(0)\rangle$ and
$\tau_{0}$ is an arbitrary time for the use of the boundary conditions. The final state will be $|\psi(T)\rangle=\sum_{n}c_{n}|\phi_{n}(\tau_{f})\rangle$,
where $\tau_{f}$ is also a time for setting the boundary conditions.
According to Sec. II, the three moving states in the simplest form are $|\phi_{1}(t)\rangle=\cos{\theta}|1\rangle+\sin{\theta}|3\rangle$,
$|\phi_{2}(t)\rangle=\sin{\theta}\cos{\varphi}|1\rangle+\sin{\varphi}|2\rangle-\cos{\theta}\cos{\varphi}|3\rangle$,
and $|\phi_{3}(t)\rangle=\sin{\theta}\sin{\varphi}|1\rangle-\cos{\varphi}|2\rangle-\cos{\theta}\sin{\varphi}|3\rangle$.
Supposing the initial state is $|1\rangle$ and the target state is
$|\psi(T)\rangle=\mu|1\rangle+\eta|2\rangle+\nu|3\rangle$ ($\mu^2+\eta^2+\nu^2=1$).
According to the initial condition, we have
\begin{eqnarray}\label{eq4-1}
  c_{1}&=&\langle\phi_{1}(\tau_{0})|1\rangle=\cos{\theta_{0}}, \cr\cr
  c_{2}&=&\langle\phi_{2}(\tau_{0})|1\rangle=\sin{\theta_{0}}\cos{\varphi_{0}}, \cr\cr
  c_{3}&=&\langle\phi_{3}(\tau_{0})|1\rangle=\sin{\theta_{0}}\sin{\varphi_{0}},
\end{eqnarray}
and according to the final condition, we have
\begin{eqnarray}\label{eq4-2}
  c_{1}&=&\langle\phi_{1}(\tau_{f})|\psi(T)\rangle \cr\cr
       &=&\mu\cos{\theta_{f}}+\nu\sin{\theta_{f}}, \cr\cr
  c_{2}&=&\langle\phi_{2}(\tau_{f})|\psi(T)\rangle \cr\cr
       &=&\mu\sin{\theta_{f}}\cos{\varphi_{f}}+\eta\sin{\varphi_{f}}-\nu\cos{\theta_{f}}\cos{\varphi_{f}}, \cr\cr
  c_{3}&=&\langle\phi_{3}(\tau_{f})|\psi(T)\rangle  \cr\cr
       &=&\mu\sin{\theta_{f}}\sin{\varphi_{f}}-\eta\cos{\varphi_{f}}-\nu\cos{\theta_{f}}\sin{\varphi_{f}},
\end{eqnarray}
where $\theta_{0}=\theta(\tau_{0})$, $\theta_{f}=\theta(\tau_{f})$, $\varphi_{0}=\varphi(\tau_{0})$, and $\varphi_{f}=\varphi(\tau_{f})$.
A set of equations are obtained
\begin{eqnarray}\label{eq4-3}
  \cos{\theta_{0}}&=&\mu\cos{\theta_{f}}+\nu\sin{\theta_{f}}, \cr\cr
  \sin{\theta_{0}}\cos{\varphi_{0}}&=&\mu\sin{\theta_{f}}\cos{\varphi_{f}}+\eta\sin{\varphi_{f}}-\nu\cos{\theta_{f}}\cos{\varphi_{f}}, \cr\cr
  \sin{\theta_{0}}\sin{\varphi_{0}}&=&\mu\sin{\theta_{f}}\sin{\varphi_{f}}-\eta\cos{\varphi_{f}}-\nu\cos{\theta_{f}}\sin{\varphi_{f}}.
\end{eqnarray}

Obviously, there are four unknowns $\theta_{0}=\theta(\tau_{0})$, $\theta_{f}=\theta(\tau_{f})$, $\varphi_{0}=\varphi(\tau_{0})$,  and $\varphi_{f}=\varphi(\tau_{f})$
in a set of three equations.
So, in order to solve the equations set in eq. (\ref{eq4-3}), it is better to confirm one of the unknowns.
For example, we can set $\varphi_{0}=0$ to make $c_{3}=0$
such that the time evolution of the system is irrelevant to the moving state $|\phi_{3}(t)\rangle$.
In this case, the results of eq. (\ref{eq4-3}) are
\begin{eqnarray}\label{eq4-4}
  \cos{\theta_{0}}&=&\mu\cos{\theta_{f}}+\nu\sin{\theta_{f}}, \cr\cr
  \sin{\theta_{0}}\sin{\varphi_{f}}&=&\eta, \cr\cr
  \sin{\theta_{0}}\cos{\varphi_{f}}&=&\mu\sin{\theta_{f}}-\nu\cos{\theta_{f}}.
\end{eqnarray}
Here, we should notice that since $\sin{\varphi_{f}}=\sin{(\pi-\varphi_{f})}$,
according to the second equation of eq (\ref{eq4-4}),
there would be two results
for $\varphi_{f}$, say $\varphi_{f}=\arcsin{(\eta/\sin{\theta_{0}})}$ and $\varphi_{f}=\pi-\arcsin{(\eta/\sin{\theta_{0}})}$.
So, it is better to use the third equation in eq. (\ref{eq4-4}) to determine $\varphi_{f}$,
then, go back and check out which one of the two results is correct.
As we have mentioned in Sec. III B that a protocol would be relatively easy to be realized if the microwave field is not needed.
In view of that, according to eq. (\ref{eq4-4}), we can further set $\theta_{0}=\theta(t)=\theta_{f}=$const to make $\dot{\theta}=0$.
We obtain
\begin{eqnarray}\label{eq4-5a}
  \theta_{0}&=&\theta_{f}=\arctan{(\frac{1-\mu}{\nu})},  \cr\cr
  \varphi_{f}&=&\arccos{(\mu-\nu\cot{\theta_{0}})}.
\end{eqnarray}
That is, the boundary condition for the system is confirmed
\begin{eqnarray}\label{eq4-5b}
  \dot{\varphi}_{0}&=&0,\ \dot{\varphi}_{f}=0,  \cr\cr
  \varphi_{0}&=&0,\ \varphi_{f}=\arccos{(\mu-\nu\cot{\theta_{0}})}.
\end{eqnarray}
Also, by fitting of a three-order polynomial $\varphi=\sum_{j=0}^{3}a_{a}t^{j}$,
we have
\begin{eqnarray}\label{eq4-5c}
  \varphi&=&\zeta[\frac{3(t-\tau_{0})^{2}}{(\tau_{f}-\tau_{0})^{2}}-\frac{2(t-\tau_{0})^{3}}{(\tau_{f}-\tau_{0})^{3}}],    \cr\cr
  \dot{\varphi}&=&6\zeta[\frac{t-\tau_{0}}{(\tau_{f}-\tau_{0})^{2}}-\frac{(t-\tau_{0})^{2}}{(\tau_{f}-\tau_{0})^{3}}],
\end{eqnarray}
where $\zeta=\arccos{(\mu-\nu\cot{\theta_{0}})}=\varphi_{f}$,
and the Rabi frequencies are
\begin{eqnarray}\label{eq4-5d}
  \Omega_{p}&=&6\zeta[\frac{t-\tau_{0}}{(\tau_{f}-\tau_{0})^{2}}-\frac{(t-\tau_{0})^{2}}{(\tau_{f}-\tau_{0})^{3}}]\sin{\theta_{0}},\cr\cr
  \Omega_{s}&=&6\zeta[\frac{t-\tau_{0}}{(\tau_{f}-\tau_{0})^{2}}-\frac{(t-\tau_{0})^{2}}{(\tau_{f}-\tau_{0})^{3}}]\cos{\theta_{0}}.
\end{eqnarray}

We plot Fig. 8 to show time-evolutions for pulses [Fig.
8 (a)] and corresponding populations transfer [Fig.
8 (b)]. We take
$|\psi(T)\rangle=\frac{1}{\sqrt{3}}(|1\rangle+|2\rangle+|3\rangle)$
as the target state, which corresponds to
$\mu=\eta=\nu={1}/{\sqrt{3}}$. By comparison with the single-mode
driving in protocol II, the benefit of multi-mode driving is clear,
the initial state is not necessary to be prepared in $|2\rangle$,
the designed process will guide the system from an arbitrary initial
state to the target state without using the microwave field. To
demonstrate that arbitrary target state would be achieved by using
multi-mode driving, in Fig. 9, we plot time-evolution for
the populations for states $|1\rangle$, $|2\rangle$, and $|3\rangle$
when : (a) $\mu=\eta=0$, $\nu=1$; (b) $\mu=0$,
$\eta=\nu=1/\sqrt{2}$; (c) $\mu=1/\sqrt{2}$, $\eta=\nu=1/2$; (d)
$\mu=1/\sqrt{2}$, $\eta=0$, $\nu=1/\sqrt{2}$. The results show that,
as expected, the multi-mode driving would guide the system to attain
arbitrary target state in an ideal way.


Noting that energy consumption is also an important index for the effectiveness of the protocol, we
calculate the behavior of the time-averaged
frequency (interpreted geometrically as a ¡°length¡± in \cite{Jmp432017}),
\begin{eqnarray}\label{eq4-5f}
  \bar{\Omega}\equiv \frac{1}{T}\int_{\tau_{0}}^{\tau_{f}}\sqrt{\Omega_{p}^{2}+\Omega_{s}^{2}}dt,
\end{eqnarray}
and the time-averaged energy
\begin{eqnarray}\label{eq4-5e}
  \bar{E}=\hbar\int_{\tau_{0}}^{\tau_{f}}{(\Omega_{p}^{2}+\Omega_{s}^{2})}dt,
\end{eqnarray}
where $T=\tau_{f}-\tau_{0}$. For comparison, we put eqs. (\ref{eq2b-7}) and (\ref{eq4-5d}) into eqs. (\ref{eq4-5f}) and (\ref{eq4-5e}).
The results are
\begin{eqnarray}\label{eq4-5g}
  \frac{\bar{\Omega}_{o}}{\bar{\Omega}_{m}}=|\frac{\arcsin{\eta}-\frac{\pi}{2}}{\arccos(\mu-\nu\cot{\theta_{0}})}|
                                           =|\frac{\arcsin{\eta}-\frac{\pi}{2}}{\pi-\arcsin({\eta}/{\sin{\theta_{0}}})}|,
\end{eqnarray}
and
\begin{eqnarray}\label{eq4-5h}
  \frac{\bar{E}_{o}}{\bar{E}_{m}}=\frac{(\arcsin{\eta}-\frac{\pi}{2})^{2}}{[\arccos(\mu-\nu\cot{\theta_{0}})]^{2}}
                                           =\frac{(\arcsin{\eta}-\frac{\pi}{2})^{2}}{[\pi-\arcsin({\eta}/{\sin{\theta_{0}}})]^{2}},
\end{eqnarray}
where the subscripts $o$ and $m$ denote the single-mode driving protocol and the multi-mode driving protocol, respectively.
Obviously, as $\eta$ ranges from $0$ to $1$, the ratio $\frac{\bar{\Omega}_{o}}{\bar{\Omega}_{m}}$ or $\frac{\bar{E}_{o}}{\bar{E}_{m}}$ is less than $1$.
Figure 10 which shows the ratio $\frac{\bar{\Omega}_{o}}{\bar{\Omega}_{m}}$ versus $\eta$ and $\mu$ also demonstrates this point.
This result signifies that the multi-mode driving protocol consumes more energy than the single-mode driving protocols.

\section*{Example for coefficients containing phases}
The discussions above have not considered the situation when the moving states' coefficients contain phases.
However, it is possible that one may ask for a target state in form of
\begin{eqnarray}\label{eq6-1}
  |\psi\rangle=|\mu||1\rangle+e^{i\gamma}|\eta||2\rangle+e^{i\kappa}|\nu||3\rangle.
\end{eqnarray}
Therefore, in this section, we will discuss how to obtain such a target state.
Obviously, to obtain a target state like that in eq. (\ref{eq6-1}), the moving state should be in form
\begin{eqnarray}\label{eq6-2}
  |\phi_{1}(t)\rangle=\sin{\theta}\cos{\varphi}|1\rangle+e^{i\gamma}\sin{\varphi}|2\rangle+e^{i\kappa}\cos{\theta}\cos{\varphi}|3\rangle.
\end{eqnarray}
Generally speaking, it does not matter whether $\gamma$ and $\kappa$ are time-dependent or not,
but for the veracity of discussion, we set $\gamma$ and $\kappa$ are time-dependent in the following.
Using the analysis above in Sec. III and Sec. IV, we can easily set the other two moving states to complete the system
\begin{eqnarray}\label{eq6-3}
  |\phi_{2}(t)\rangle&=&\sin{\theta}\sin{\varphi}|1\rangle-e^{i\gamma}\cos{\varphi}|2\rangle+e^{i\kappa}\cos{\theta}\sin{\varphi}|3\rangle, \cr\cr
  |\phi_{3}(t)\rangle&=&\cos{\theta}|1\rangle-e^{i\kappa}\sin{\theta}|3\rangle.
\end{eqnarray}
Then, the Hamiltonian is given as
\begin{eqnarray}\label{eq6-4}
  H&=&i\sum_{n=1}^{3}{|\dot{\phi}_{n}\rangle\langle\phi_{n}|} \cr\cr
   &=&\left(\begin{array}{ccc}
        0 & -ie^{-i\gamma}\dot{\varphi}\sin{\theta} & ie^{-i\kappa}\dot{\theta} \\
        ie^{i\gamma}\dot{\varphi}\sin{\theta} & -\dot{\gamma} & i\dot{\varphi}e^{-i(\kappa-\gamma)}\cos{\theta} \\
        -ie^{i\kappa}\dot{\theta} & -i\dot{\varphi}e^{i(\kappa-\gamma)}\cos{\theta} & -\dot{\kappa}
      \end{array}
      \right).
\end{eqnarray}
Contrasting eq. (\ref{eq6-4}) with eq. (\ref{eq2-2}), we can find the Hamiltonians in eqs. (\ref{eq6-4}) and (\ref{eq2-2}) are very close to each other.
Therefore, we can similarly consider the system is a three-level $\Lambda$-type system.
The three Rabi frequencies are
$\Omega_{p}=\dot{\varphi}\sin{\theta}$, $\Omega_{s}=\dot{\varphi}\cos{\theta}$, and $\Omega_{a}=\dot{\theta}$.
Then, at the instance of the final state, we can suitably set the boundary conditions to design the pulses and perform
the QSE scheme as what we do in Sec. III and Sec. IV.
For instance, assuming the initial state is $|3\rangle$ and the target state is $|\mu||1\rangle+e^{i\kappa}|\nu||3\rangle$,
the coefficients could be chosen as
\begin{eqnarray}\label{eq6-5}
  \theta&=&(\frac{3t^2}{T^2}-\frac{2t^3}{T^3})\arcsin{\mu},\cr\cr
  \dot{\theta}&=&6(\frac{t}{T^{2}}-\frac{t^{2}}{T^{3}})\arcsin{\mu}, \cr\cr
  \varphi&=&0,\ \ \kappa=\lambda\pi t,
\end{eqnarray}
where $\lambda$ is a coefficient decided by the target state.
To demonstrate the time evolution governed by the designed Hamiltonian $H$ is accurately along the moving state when phases are considered,
we extract $\theta$ and $\kappa$ from $|\psi(t)\rangle$ with relations ($-\pi/2<\theta<\pi/2$)
\begin{eqnarray}\label{eq6-7}
  \theta'&=&\arcsin\sqrt{|\langle1|\psi(t)\rangle|^{2}}, \cr\cr
  \kappa'&=&-i\ln\left[{\frac{\langle3|\psi(t)\rangle}{|\langle3|\psi(t)\rangle|}}\right],
\end{eqnarray}
where $|\psi(t)\rangle$ is the solution of Schr\"{o}dinger
equation $i\partial_{t}|\psi(t)\rangle=H|\psi(t)\rangle$.
Based on eqs. (\ref{eq6-5}) and (\ref{eq6-7}), we plot $\theta$, $\kappa$, $\theta'$, and $\kappa'$ versus time in Fig. 11 with $\mu=1/\sqrt{2}$.
Obviously, from the figure, we find $\theta=\theta'$ and $\kappa=\kappa'$.
That is, when the phases are considered, the system still evolves along the way as we expected.
In fact, when $\varphi=0$, the system can be regarded as a two-state system,
in which any quantum state can be depicted as a point at the origin of the Bloch sphere.
We plot the time evolution of the system in Fig. 12 with
the help of Bloch sphere when $r=1$, $\mu=1/\sqrt{2}$, and $\lambda=0.5/T$.
Any point on the black solid curve expresses a quantum
state containing information of population and phase.
As the information of phase during the whole evolution is accurately known,
the present QSE scheme has a good application prospect in quantum phase gates.

\section*{Discussion and Conclusion}
An classical application of the QSE scheme could be the implementation of beam splitters
in a system with longitudinal coordinate three coupled waveguides \cite{Lpr3243,Jpb44051001,Jpb41085402,Pra85055803,Pra89053408}, or in a system
with a single particle in a triple well \cite{Pra70023606}.
For such systems the minimal
channel basis for left, center, and right wave functions are
$|C\rangle=|2\rangle=[0,1,0]^{t}$, $|L\rangle=|1\rangle=[1,0,0]^{t}$, and $|R\rangle=|3\rangle=[0,0,1]^{t}$, where
the superscript $t$ means transposition.
The Rabi frequencies $\Omega_{p}$, $\Omega_{s}$, and $\Omega_{a}$ play the roles of coupling coefficients
between the adjacent waveguides or between the adjacent wells.
For example, if we would like to implement a $1:2$ beam splitter, the goal is to drive the system from $|C\rangle$ to $\frac{1}{\sqrt{2}}(|L\rangle+|R\rangle)$.
So, we can choice protocol II to realize the process.
The boundary condition is given according to eq. (\ref{eq2b-5}), and the coefficients for the final state are $\{\eta=0,\ \mu=\nu=\frac{1}{\sqrt{2}}\}$.
If the coefficients for the final state are chosen as $\{\eta=\mu=\nu=\frac{1}{\sqrt{3}}\}$, the protocol II can be used to implement a $1:3$ beam splitter.
The QSE scheme also has a good application prospect in the field of multiparticle entanglement generation, for instance, by Rydberg blockade \cite{Prl100170504,Njp16042025}.
For the multi-mode driving protocol, the application area would be much wider as the
Hamiltonian turns out to be a stimulated Raman passage Hamiltonian,
for example, the multi-mode driving protocol would be applied in field of entangled states' fast generation in cavity quantum electrodynamics (QED) systems \cite{CYH}.
As it is known to all, the 1-3 pulse has been regarded as a problematic term \cite{Pra87043402,Pra89043408,Pra89053408,CYH} because it is possibly an outstanding challenge to realize
the 1-3 pulse in some specific systems.
Therefore, in the schemes proposed in ref. \cite{CYH}, the authors did a lot to design Hamiltonian to
overcome the problem caused by the problematic term which is actually equivalent to the special one-photon 1-3 pulse (the microwave field).
However, the operations in ref. \cite{CYH} usually cause other problem or make other limiting conditions to the schemes, for examples, there
will be a limiting condition for the total operation time to generate the entangled states.
So, researchers never ceased finding realizable methods to replace or nullify the problematic term in some specific systems.
In this paper, we find that by applying the multi-mode driving, the problem will probably be avoided because the multi-mode driving
allows the designed Hamiltonian without using the problematic terms to guide the system to achieve the target state.
Taking a cavity QED system with two two-level atoms (ground state $|g\rangle$ and excited state $|e\rangle$)
in a cavity as an example, the Hamiltonian for the one-excited subspace under rotating wave approximation is ($\hbar=1$)
\begin{eqnarray}\label{eq5-1}
  H=\left(
      \begin{array}{ccc}
        0 & g_{1} & 0 \\
        g_{1}^{*} & 0 & g_{2}^{*} \\
        0 & g_{2} & 0
      \end{array}
    \right),
\end{eqnarray}
where the basis for the Hamiltonian are $|1\rangle=|e,g\rangle_{12}|0\rangle_{c}$, $|2\rangle=|g,g\rangle_{12}|1\rangle_{c}$,
and $|3\rangle=|g,e\rangle_{12}|0\rangle_{c}$.
We just need to set $g_{1}=-i\Omega_{p}$ and $g_{2}=i\Omega_{s}$ according to eq. (\ref{eq4-5d}).
Choosing $\mu=\eta=1/\sqrt{2}$, time-evolution of the system [see Fig. 13] will end up with $|\psi(T)\rangle=(|e,g\rangle_{12}+|g,e\rangle_{12})/\sqrt{2}$ which
is a two-atom maximal entangled state.

Moreover, this universal QSE scheme can be extended straightforwardly
into higher-dimensional systems.
For example, a simple set of moving states for a four-dimensional system could be given as
$|\phi_{n}(t)\rangle=\frac{1}{\sqrt{2}}(\cos{\alpha_{n}}\cos{\beta_{n}}|1\rangle+\cos{\alpha_{n}}\sin{\beta_{n}}|2\rangle
+\sin{\alpha_{n}}\cos{\beta_{n}}|3\rangle+\sin{\alpha_{n}}\sin{\beta_{n}}|4\rangle)$.
Then, the orthogonality condition requests $\cos{(\alpha_{n}-\alpha_{m})\cdot\cos{(\beta_{n}-\beta_{m})}}=0$.
We can accordingly set the four moving states as
\begin{eqnarray}\label{eq4-5}
  |\phi_{1}\rangle&=&\frac{1}{\sqrt{2}}(\cos{\theta}\cos{\varphi}|1\rangle+\cos{\theta}\sin{\varphi}|2\rangle   \cr\cr
                                       &&+\sin{\theta}\cos{\varphi}|3\rangle+\sin{\theta}\sin{\varphi}|4\rangle), \cr\cr
  |\phi_{2}\rangle&=&\frac{1}{\sqrt{2}}(\sin{\theta}\cos{\varphi}|1\rangle+\sin{\theta}\sin{\varphi}|2\rangle   \cr\cr
                                       &&-\cos{\theta}\cos{\varphi}|3\rangle-\cos{\theta}\sin{\varphi}|4\rangle), \cr\cr
  |\phi_{3}\rangle&=&\frac{1}{\sqrt{2}}(\cos{\theta}\sin{\varphi}|1\rangle-\cos{\theta}\cos{\varphi}|2\rangle   \cr\cr
                                       &&+\sin{\theta}\sin{\varphi}|3\rangle-\sin{\theta}\cos{\varphi}|4\rangle), \cr\cr
  |\phi_{4}\rangle&=&\frac{1}{\sqrt{2}}(\sin{\theta}\sin{\varphi}|1\rangle-\sin{\theta}\cos{\varphi}|2\rangle    \cr\cr
                                       &&-\cos{\theta}\sin{\varphi}|3\rangle+\cos{\theta}\cos{\varphi}|4\rangle).
\end{eqnarray}
The corresponding Hamiltonian is
\begin{eqnarray}\label{eq4-6}
  H(t)&=&i\hbar\sum_{n}|\dot{\phi}_{n}(t)\rangle\langle\phi_{n}(t)| \cr\cr
      &=&
         \frac{\hbar}{2}\left(
           \begin{array}{cccc}
             0 & -i\dot{\varphi} & -i\dot{\theta} & 0 \\
             i\dot{\varphi} & 0 & 0 & -i\dot{\theta} \\
             i\dot{\theta} & 0 & 0 & -i\dot{\varphi} \\
             0 & i\dot{\theta} & i\dot{\varphi} & 0
           \end{array}
         \right).
\end{eqnarray}
Then, according to the initial state and the target state, we set the boundary condition to construct the pulses
to determine the Hamiltonian. By applying the constructed Hamiltonian, arbitrary quantum states in a four-dimensional system will be achieved.


In conclusion, we have proposed an effective scheme for arbitrary quantum state engineering via Counterdiabatic driving.
The scheme enables to achieve an arbitrarily fast population transfer in a three-state quantum system.
As it is known to all, the transitionless tracking algorithm provides a Hamiltonian
to accurately drive one of the eigenstates of an original Hamiltonian without transition to other ones.
Different from the previous work based on the transitionless tracking algorithm that the start
point is usually assuming $H_{0}(t)|\phi_{n}(t)\rangle=E_{n}(t)|\phi_{n}(t)\rangle$ which means
$|\phi_{n}(t)\rangle$ should be the eigenstate of $H_{0}(t)$, in this paper, we directly start from a time-dependent moving state $|\phi_{n}(t)\rangle$ to
design a process to achieve the target state. Strictly speaking, the moving states $\{|\phi_{n}(t)\rangle\}$
are probably (but not necessary) the eigenstates of an unknown Hamiltonian because $\{|\phi_{n}(t)\rangle\}$ satisfy orthonormality.
While that does not matter because the unknown Hamiltonian affects nothing to the proposed QSE scheme.
In this QSE scheme, the key point is to make sure that the moving states satisfy orthonormality which means
$\sum_{n}|\phi_{n}(t)\rangle\langle\phi_{n}(t)|=1$ and $\langle\phi_{n}(t)|\phi_{m}(t)\rangle=\delta_{nm}$.
We have proposed different protocols based on single-mode driving and multi-mode driving as examples to discuss the QSE scheme.
The result shows that all the protocols, especially, the multi-mode driving protocol,
can realize the target state in a perfect way: guiding the system to attain an arbitrary
target state at a predefined time. The only drawback is that by single-mode driving,
there might be some limits for the initial condition according to some special requirements.
For example, if it is impossible to perform the one-photon 1-3 pulse (the microwave field),
in order to achieve an arbitrary target state, the
initial state should be ideally in the intermediate state, i.e., $|2\rangle$.
Moreover, the pulses designed by polynomial fitting to realize the
QSE scheme are shown as the shapes of sinusoid or linear superposition of sinusoids,
which means realizing the QSE scheme in practice is not a challenge.
Therefore, we hope the QSE scheme would be possible to realize within the current
experimental technology.

\section*{Acknowledgement}

  This work was supported by the National Natural Science Foundation of China under Grants No.
11575045 and No. 11374054, the Foundation of Ministry of Education
of China under Grant No. 212085, and the Major State Basic Research
Development Program of China under Grant No. 2012CB921601.

\section*{Author Contributions}

Y. H. C and Y. X. came up with the initial idea for the work and
performed the simulations for the model. J. S. performed the
calculations for the model.  B. H. H, Q. C. W, and Y. H. C. performed all the
data analysis and the initial draft of the manuscript. All authors
participated in the writing and revising of the text.

\section*{Additional Information}

Competing financial interests: The authors
declare no competing financial interests.

\newpage

\begin{figure}
 \scalebox{0.3}{\includegraphics {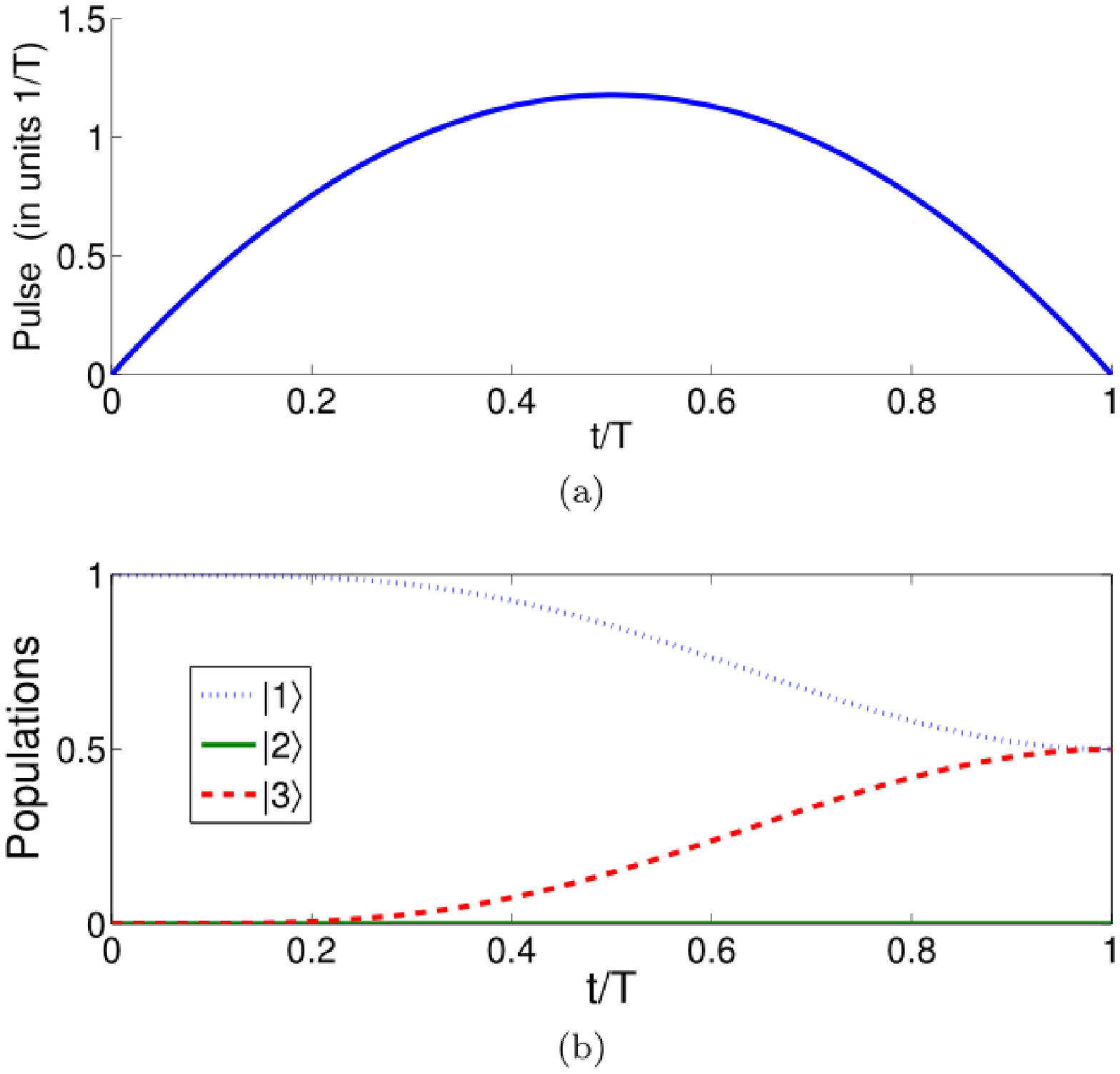}}
 \caption{
          Protocol I of single-mode driving:
    (a) Dependence on $t/T$ of the Rabi frequency $\Omega_{a}(t)$ when $\mu={1}/{\sqrt{2}}$ .
    (b) Time-evolution for states $|1\rangle$, $|2\rangle$, and $|3\rangle$ when $\nu={1}/{\sqrt{2}}$.
         }
 \label{fig1}
\end{figure}
\begin{figure}
 \scalebox{0.3}{\includegraphics {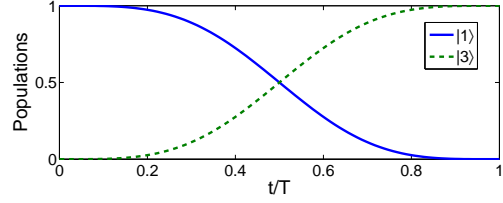}}
 \caption{
          Protocol I of single-mode driving: Time-evolution populations for states $|1\rangle$ and $|3\rangle$ when $\nu=1$ for the arbitrarily fast population inversion.
         }
 \label{fig2}
\end{figure}
\begin{figure}
 \scalebox{0.3}{\includegraphics {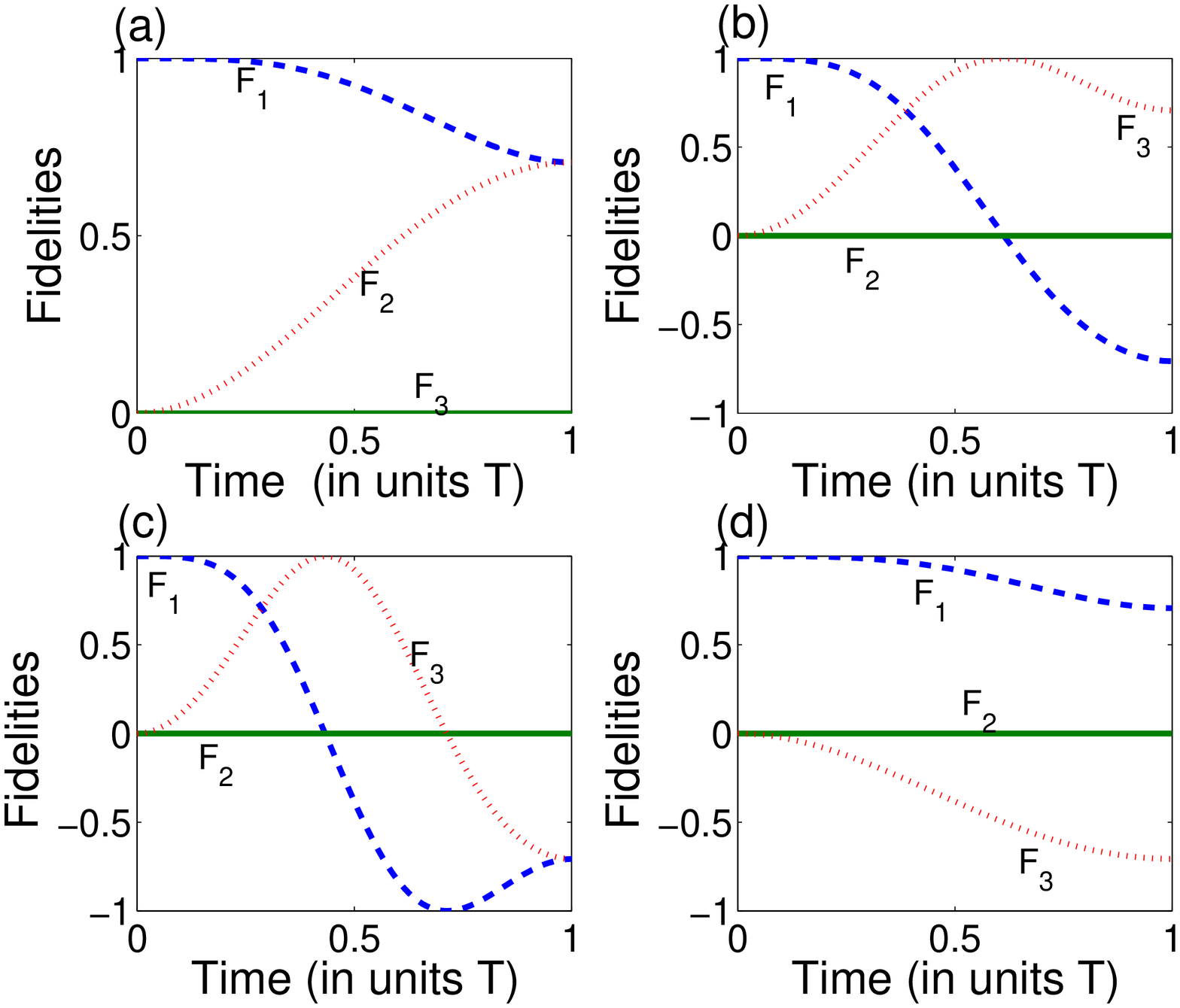}}
 \caption{
          Protocol I of single-mode driving: The fidelities for states $|1\rangle$, $|2\rangle$, and $|3\rangle$ with $|\mu|=|\nu|=1/\sqrt{2}$ when:
          (a) $\theta(T)=\arccos{|\mu|}=\arcsin{|\nu|}$;
          (b) $\theta(T)=\pi-\arccos{|\mu|}$;
          (c) $\theta(T)=\pi+\arccos{|\mu|}=\pi+\arcsin{|\nu|}$;
          (d) $\theta(T)=-\arcsin{|\nu|}$.
         }
 \label{fig3}
\end{figure}
\begin{figure}
 \scalebox{0.3}{\includegraphics {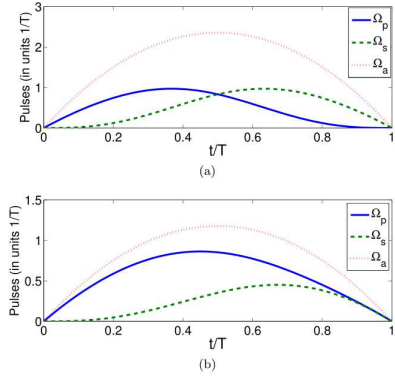}}
 \caption{
          Protocol II of single-mode driving: Rabi frequencies $\Omega_{p}(t)$, $\Omega_{s}(t)$, and $\Omega_{a}(t)$ when
         (a) $\mu=0$, $\eta=\nu={1}/{\sqrt{2}}$;
         (b) $\mu=\eta=\nu={1}/{\sqrt{3}}$.
         }
 \label{fig4}
\end{figure}
\begin{figure}
 \scalebox{0.3}{\includegraphics {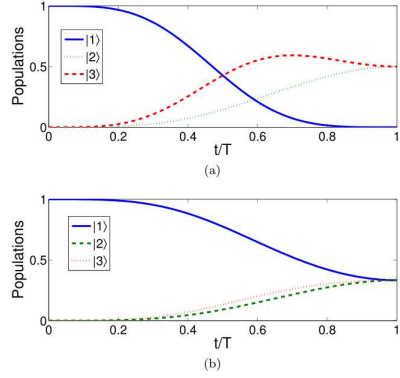}}
 \caption{
          Protocol II of single-mode driving: Time-evolution populations when
        (a) $\mu=0$, $\eta=\nu={1}/{\sqrt{2}}$;
        (b) $\mu=\eta=\nu={1}/{\sqrt{3}}$.
         }
 \label{fig5}
\end{figure}
\begin{figure}
 \scalebox{0.3}{\includegraphics {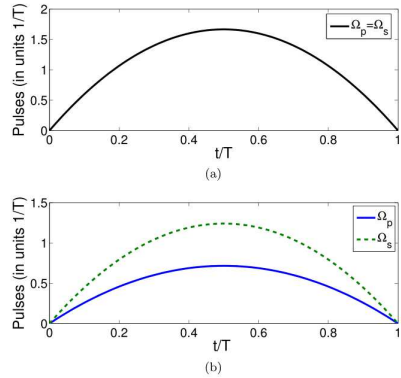}}
 \caption{
         Protocol II of single-mode driving: Dependence on $t/T$ of the Rabi frequencies $\Omega_{p}(t)$, $\Omega_{s}(t)$, and $\Omega_{a}(t)$ in case of $\dot{\theta}=0$
         (without using the 1-3 pulse) when
         (a) $\mu={1}/{\sqrt{2}}$, $\eta=0$, and $\nu={1}/{\sqrt{2}}$;
         (b) $\mu={1}/{\sqrt{6}}$, $\eta={1}/{\sqrt{3}}$, and $\nu={1}/{\sqrt{2}}$.
         }
 \label{fig6}
\end{figure}
\begin{figure}
 \scalebox{0.3}{\includegraphics {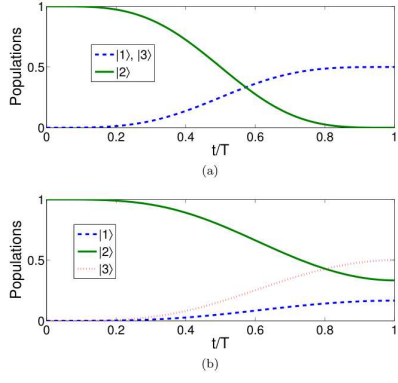}}
 \caption{
         Protocol II of single-mode driving: Time-evolution populations for states $|1\rangle$, $|2\rangle$, and $|3\rangle$ in case of $\dot{\theta}=0$ when
        (a) $\mu={1}/{\sqrt{2}}$, $\eta=0$, and $\nu={1}/{\sqrt{2}}$;
        (b) $\mu={1}/{\sqrt{6}}$, $\eta={1}/{\sqrt{3}}$, and $\nu={1}/{\sqrt{2}}$.
         }
 \label{fig7}
\end{figure}
\begin{figure}
 \scalebox{0.3}{\includegraphics {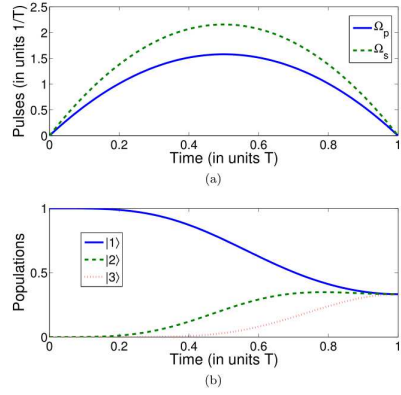}}
 \caption{
         Protocol of multi-mode driving:
        (a) Dependence on $t/T$ of the Rabi frequencies $\Omega_{p}(t)$ and $\Omega_{s}$ when $\mu=\eta=\nu={1}/{\sqrt{3}}$.
        (b) Time-evolution for states $|1\rangle$, $|2\rangle$, and $|3\rangle$ when $\mu=\eta=\nu={1}/{\sqrt{3}}$.
         }
 \label{fig8}
\end{figure}
\begin{figure}
 \scalebox{0.3}{\includegraphics {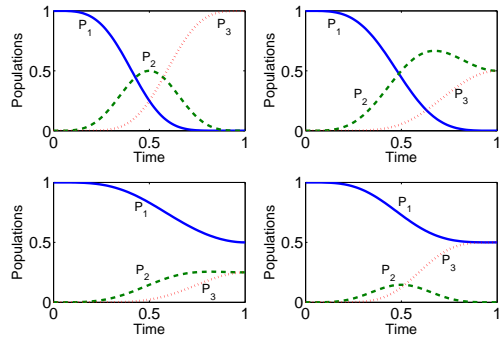}}
 \caption{
         Protocol of multi-mode driving:
         Time-evolution populations for states $|1\rangle$, $|2\rangle$, and $|3\rangle$ when
        (a) $\mu=\eta=0$, $\nu=1$;
        (b) $\mu=0$, $\eta=\nu=1/\sqrt{2}$;
        (c) $\mu=1/\sqrt{2}$, $\eta=\nu=1/2$;
        (d) $\mu=1/\sqrt{2}$, $\eta=0$, $\nu=1/\sqrt{2}$.
         }
 \label{fig9}
\end{figure}
\begin{figure}
 \scalebox{0.3}{\includegraphics {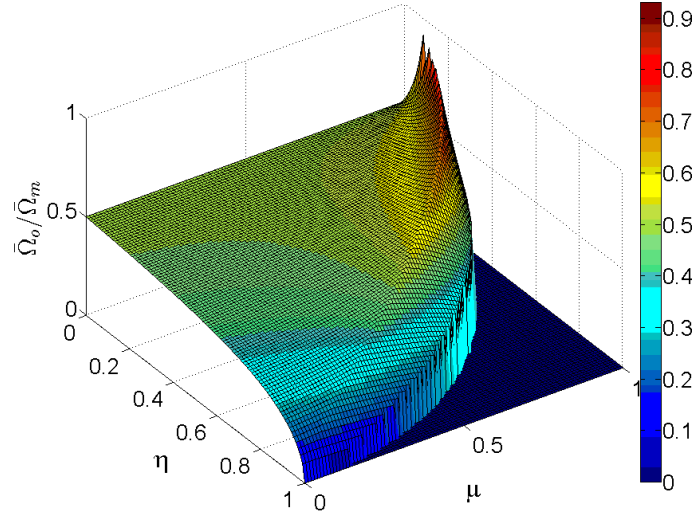}}
 \caption{
         The ratio ${\bar{\Omega}_{o}}/{\bar{\Omega}_{m}}$ versus $\eta$ and $\mu$,
         we impose ${\bar{\Omega}_{o}}/{\bar{\Omega}_{m}}=0$ when $\eta^2+\mu^2>1$ in plotting the figure.
         }
 \label{fig10}
\end{figure}
\begin{figure}
 \scalebox{0.3}{\includegraphics {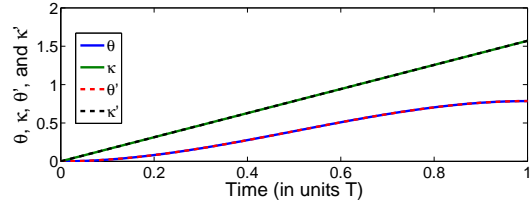}}
 \caption{
         Time-evolution for the system on the Bloch sphere when $\mu=1/\sqrt{2}$ and $\lambda=0.5/T$.
         }
 \label{fig11}
\end{figure}
\begin{figure}
 \scalebox{0.3}{\includegraphics {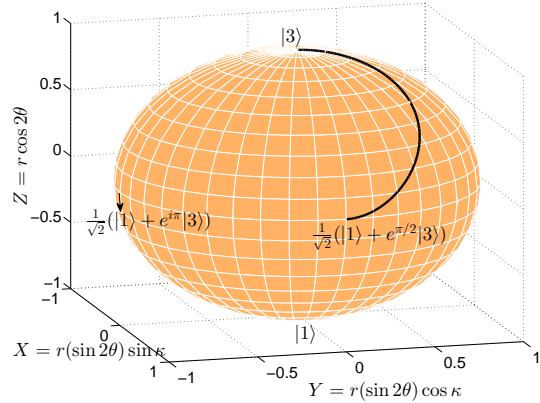}}
 \caption{
         Time-evolution for $\theta$, $\kappa$, $\theta'$, and $\kappa'$ when $\mu=1/\sqrt{2}$ and $\lambda=0.5/T$.
         }
 \label{fig12}
\end{figure}
\begin{figure}
 \scalebox{0.3}{\includegraphics {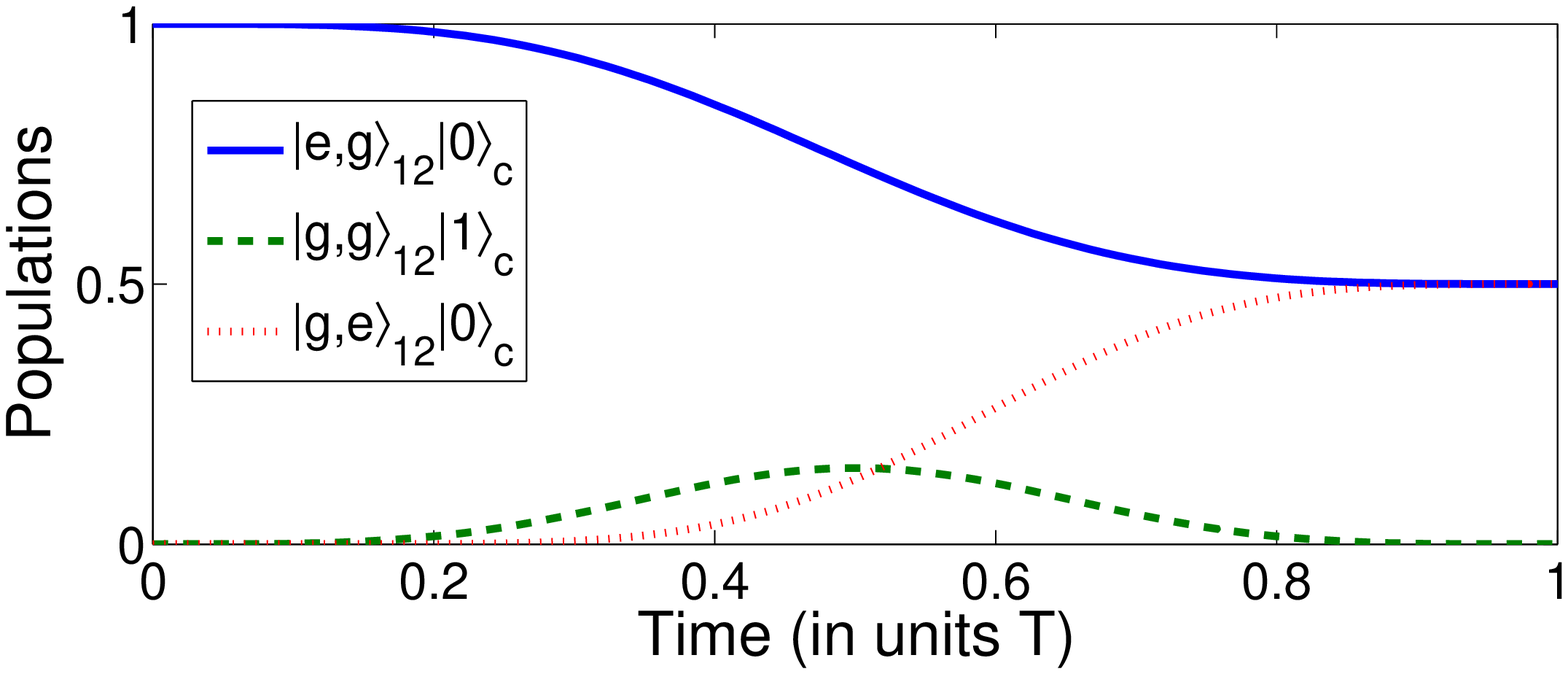}}
 \caption{
         Time-evolution for states $|e,g\rangle_{12}|0\rangle$, $|g,g\rangle_{12}|1\rangle$, and $|g,e\rangle_{12}|0\rangle$ for the example of entangled state's generation.
         }
 \label{fig13}
\end{figure}

\end{document}